\documentclass[12pt,a4paper,reqno]{article}
\usepackage{amsmath}

\advance\textheight\topskip
\allowdisplaybreaks

\newcommand{\w}{\omega}
\newcommand{\hw}{\hat{\omega}}

\newcommand{\e}{\mathrm{e}}

\newcommand{\ep}{\varepsilon}
\newcommand{\tab}{\quad\,}
\newcommand{\p}{\partial}
\newcommand{\cD}{\mathcal{D}}
\newcommand{\cE}{\mathcal{E}}
\newcommand{\cF}{\mathcal{F}}

\newcommand{\half}{\frac{\raisebox{-2pt}{$1$}}{2}}
\newcommand{\frc}[2]{\frac{\raisebox{-2pt}{$#1$}}{#2}}

\DeclareMathOperator{\dual}{\star}
\DeclareMathOperator{\hstar}{\hat{\star}}

\begin{document}

\begin{flushright} \small
 ITP-UH-19/99 \\ UB-ECM-PF-99/17 \\ hep-th/9910177
\end{flushright}
\medskip

\begin{center}
 {\large\bfseries Exotic Gauge Theories from Tensor Calculus}
 \\[5mm]
 Friedemann Brandt ${}^a$, Joan Sim\'on ${}^b$, Ulrich Theis ${}^a$
 \\[2mm]
 {\small\slshape
 ${}^a$\ Institut f\"ur Theoretische Physik, Universit\"at Hannover, \\
 Appelstra\ss{}e 2, 30167 Hannover, Germany \\[2mm]
 ${}^b$\ Departament ECM, Facultat de F\'{\i}sica, Universitat de
 Barcelona and \\
 Institut de F\'{\i}sica d'Altes Energies, Diagonal 647, E-08028,
 Barcelona, Spain}
\end{center}
\vspace{6mm}

\centerline{\bfseries Abstract} \medskip
We construct non-standard interactions between exterior form gauge
fields by gauging a particular global symmetry of the Einstein-Maxwell
action for such fields. Furthermore we discuss generalizations of such
interactions by adding couplings to gravitational Chern-Simons forms and
to fields arising through dimensional reduction. The construction uses
an appropriate tensor calculus.

\section*{Introduction} 

Exterior form gauge fields $A_p=(1/p!)\, dx^{\mu_1}\!\dots dx^{\mu_p}
A_{\mu_1\dots\mu_p}$ generalize naturally the electromagnetic
gauge potential and are therefore interesting on general grounds in the
context of gauge theories. In particular they play a substantial
r\^{o}le in supergravity models and string theory. It is therefore
natural to study consistent interactions of such gauge fields, both the
interactions between themselves and their couplings to other fields.

In this paper we construct non-standard interactions of exterior
form gauge fields. This is done in curved spacetime, i.e. the coupling
to the gravitational field is included as well. Our starting point is
the Einstein-Maxwell action for a $p$-form gauge field $A_p$ and an
$(n-p-1)$-form gauge field $A_{n-p-1}$ in $n$-dimensional spacetime
(for arbitrary $p$ and $n$). This action has a global symmetry which
shifts $A_p$ by the hodge dual of the field strength of $A_{n-p-1}$,
and $A_{n-p-1}$ by the hodge dual of the field strength of $A_p$
(with an appropriate sign factor, see below).  We gauge this global
symmetry, using an additional 1-form gauge field $V=dx^\mu V_\mu$.
This yields inevitably interactions which are non-polynomial in the
$V_\mu$. It would therefore be cumbersome to construct these
interactions in a pedestrian way via the standard Noether procedure.
Instead, we employ an appropriate tensor calculus which is analogous to
the one introduced in \cite{BD} in flat four-dimensional spacetime. By
formulating this tensor calculus in the differential form language we
simplify the construction considerably as compared to the formulation in
terms of components used in \cite{BD}. 

The resulting interactions can be generalized in several ways. We shall
describe two such generalizations arising in gravitational theories: (a)
additional couplings of $p$-form gauge fields to gravitational
Chern-Simons forms; (b) couplings to additional vector gauge fields and
scalar fields arising by standard Kaluza-Klein type dimensional
reduction.

Our work is linked to recent progress in four-dimensional supersymmetric
gauge theories, and theories with exterior form gauge fields in general.
To our knowledge, gauge theories of the type constructed here appeared
for the first time in \cite{CdWFKST}. There four-dimensional $N=2$
supersymmetric gauge theories were constructed in which the central
charge of the vector-tensor multiplet is gauged. This central charge is
a global symmetry of the type described above and therefore the models
found in \cite{CdWFKST} contain interactions of the same non-standard
type as those we shall obtain. Further four-dimensional $N=2$
supersymmetric gauge theories of the same type were constructed in
\cite{VT}. The vector-tensor multiplet with gauged central charged
is believed to arise in $N=2$ heterotic string vacua \cite{string}.
Interactions of the type studied here may therefore be relevant 
in that context, among other things.

Independently of these developments related to the $N=2$
vector-tensor multiplet, models of the type considered here, and
generalizations thereof, were discovered in \cite{HK} (along with
further new interactions of exterior form gauge fields) within a
systematic classification of possible consistent interaction vertices of
exterior form gauge fields. Four-dimensional $N=1$ supersymmetric
versions of some of these models were constructed in \cite{BT}. Our
work thus adds to the list of new gauge theories in these references.

\section*{The Basic Model} 

The properly normalized Einstein-Maxwell action for $A_p$ and
$A_{n-p-1}$ is
 \begin{equation}
  S_0 = \half \int \big[ (-)^{(p+1)(n-1)}\, d A_p \dual d A_p +
  (-)^{p(n-1)}\, d A_{n-p-1} \dual d A_{n-p-1} \big]\ .
 \end{equation}
Here we use the following conventions for a $p$-form $\w_p$ and its
Hodge dual:
 \begin{align*}
  \w_p & = \frc{1}{p!}\, dx^{\mu_1} \dots dx^{\mu_p}\, \w_{\mu_1 \dots
	\mu_p} \\
  \dual \w_p & = \frc{1}{p!(n-p)!}\, dx^{\mu_1} \dots dx^{\mu_{n-p}}\,
	\ep_{\mu_1 \dots \mu_n} \w^{\mu_{n-p+1} \dots \mu_n}\ ,
 \end{align*}
where indices are raised with the inverse metric $G^{\mu\nu}$, and the
curved Levi-Civita tensor is defined by ($x^1$ labels the time
coordinate)
 \begin{equation*}
  \ep_{12 \dots n} = - \sqrt{G}\ ,\quad G = - \det G_{\mu\nu}\ .
 \end{equation*}
One should of course add the Einstein action to $S_0$, which however we
shall not write explicitly.

$S_0$ has among others a global symmetry generated by
 \begin{equation} \label{DeA}
  \Delta A_p = \dual d A_{n-p-1}\ ,\quad \Delta A_{n-p-1} = -
  (-)^{n(p+1)} \dual d A_p\ .
 \end{equation}
The corresponding Noether current, written in dual notation as an
$(n-1)$-form, is
 \begin{equation} \label{current}
  J = \dual d A_p\, \dual d A_{n-p-1}\ .
 \end{equation}
We remark that the global symmetry $\Delta$ and its current $J$ are
non-trivial, see comments at the end of the paper.

We will now try to gauge this symmetry. To do this, we search for a
modified generator $\Delta'$, such that $\Delta' A_p$ and
$\Delta' A_{n-p-1}$ transform covariantly under gauge transformations
 \begin{equation} \label{deA}
  \delta_\epsilon A_p = g\, \epsilon \Delta' A_p\ ,\quad
  \delta_\epsilon A_{n-p-1} = g\, \epsilon \Delta' A_{n-p-1}\ ,
 \end{equation}
where $\epsilon$ is an arbitrary scalar field and $g$ a coupling
constant of mass dimension $-1$.

A reasonable ansatz is to replace the exterior derivative in \eqref{DeA}
with a covariant one. Let us therefore introduce a connection 1-form $V$
with the standard transformation law
 \begin{equation} \label{deV}
  \delta_\epsilon V = d \epsilon
 \end{equation}
and a covariant derivative
 \begin{equation}
  D = d - g V \Delta'\ ,
 \end{equation}
where $\Delta'$ is the covariant version of \eqref{DeA},
 \begin{equation}
  \Delta' A_p = \dual D A_{n-p-1}\ ,\quad \Delta' A_{n-p-1} = -
  (-)^{n(p+1)} \dual D A_p\ .
 \end{equation}
These equations give the action of $\Delta'$ only implicitly because
$D$ involves $\Delta'$. We now determine the covariant derivatives
$DA_p$ and $DA_{n-p-1}$. Starting with the former, we have
 \begin{align*}
  D A_p & = d A_p - g V \dual D A_{n-p-1} \\
  & = d A_p - g V \dual \big[ d A_{n-p-1} + (-)^{n(p+1)} g\, V \dual
	D A_p \big]\ .
 \end{align*}
Using the identity
 \begin{equation}
  \dual(V \dual \w_p) = (-)^{np}\, i_V \w_p\ ,\quad i_V = V^\mu\,
  \frc{\p}{\p(dx^\mu)}\ ,
 \end{equation}
which holds for any $p$-form $\w_p$ (where $V^\mu = G^{\mu\nu} V_\nu$),
this gives
 \begin{equation*}
  D A_p = d A_p - g V \dual d A_{n-p-1} - g^2 V i_V D A_p\ .
 \end{equation*}
To solve for $DA_p$, we have to invert the operator $1 + g^2 V i_V$.
With $(V i_V)^2 = V^\mu V_\mu\, V i_V$, it is easily verified that
 \begin{equation}
  (1 + g^2 V i_V)^{-1} = 1 - g^2 E^{-1} V i_V\ ,\quad
  E = 1 + g^2 V^\mu V_\mu\ ,
 \end{equation}
and we obtain
 \begin{equation} \label{DAp}
  D A_p = d A_p - g E^{-1} V \big[ \dual d A_{n-p-1} + g\, i_V d A_p
  \big]\ .
 \end{equation}
Analogously, one finds
 \begin{equation} \label{DAn-p-1}
  D A_{n-p-1} = d A_{n-p-1} - g E^{-1} V \big[ -(-)^{n(p+1)} \dual
  d A_p + g\, i_V d A_{n-p-1} \big]\ .
 \end{equation}
Note that due to the appearance of $E^{-1}$ the covariant derivatives,
and thus the gauge transformations, of $A_p$ and $A_{n-p-1}$ are
non-polynomial in the connection $V_\mu$ and the coupling constant $g$.

As can be checked, $DA_p$ and $DA_{n-p-1}$ are indeed covariant, i.e.\
their gauge transformations do not involve derivatives of $\epsilon$,
and one has
 \begin{align}
  \delta_\epsilon\, D A_p & = g\, \epsilon D \Delta' A_p = g\, \epsilon
	D \dual D A_{n-p-1} \notag \\
  \delta_\epsilon\, D A_{n-p-1} & = g\, \epsilon D \Delta' A_{n-p-1} =
	- (-)^{n(p+1)} g\, \epsilon D \dual D A_p\ . \label{deDA}
 \end{align}

Now we can proceed to construct the gauge invariant action. To do
this, we use the following fact: let $X$ be a covariant volume form
which transforms according to $\delta_\epsilon X = g\, \epsilon DK$,
with $K$ a covariant $(n-1)$-form. Then $X + g\, VK$ transforms into
a total derivative,
 \begin{equation}
  \delta_\epsilon X = g\, \epsilon D K\, ,\quad \delta_\epsilon K =
  g\, \epsilon \Delta' K \quad\Rightarrow\quad \delta_\epsilon (X +
  g\, V K) = d (g\, \epsilon K)\ .
 \end{equation}
In particular, thanks to \eqref{deDA} this applies to
 \begin{equation} \label{X}
  X = (-)^{(p+1)(n-1)}\, D A_p \dual D A_p + (-)^{p(n-1)}\, D
  A_{n-p-1} \dual D A_{n-p-1}
 \end{equation}
with
 \begin{equation} \label{K}
  K = - 2 (-)^{n-p} \dual D A_p\, \dual D A_{n-p-1}\ .
 \end{equation}
Adding a kinetic term for the connection $V$, the action thus reads
 \begin{equation} \label{S}
  S = \half \int \big( X + g\, V K + d V \dual d V \big)\ ,
 \end{equation}
with $X$ and $K$ as in \eqref{X} and \eqref{K}. By construction,
$\delta_\epsilon S$ is a surface term, with the transformations as in
\eqref{deA}, \eqref{deV} and $\delta_\epsilon G_{\mu\nu} = 0$.
Furthermore, the action is invariant under the standard spacetime
diffeomorphisms and the abelian gauge transformations of $A_p$ and
$A_{n-p-1}$,
 \begin{equation} \label{trafo1}
  \delta A_p = d \Lambda_{p-1}\ ,\quad \delta A_{n-p-1} = d
  \Lambda_{n-p-2}\ ,
 \end{equation}
since these fields enter the covariant derivatives \eqref{DAp},
\eqref{DAn-p-1} only via their exterior derivatives.
\medskip

Alternatively, one may consider a first order formulation. Introducing
auxiliary fields $\beta_p$ and $\beta_{n-p-1}$, with the form degree as
indicated, the action is given by
 \begin{align}
  S = \half \int \big[ & d V \dual d V + (-)^{p(n-1)}\,	\beta_p \dual
	\beta_p + (-)^{(p+1)(n-1)}\, \beta_{n-p-1} \dual \beta_{n-p-1}
	\notag \\[-2pt]
  & + 2 \beta_p\, d A_{n-p-1} - 2 (-)^{n(p+1)} \beta_{n-p-1}\, d A_p -
	2 (-)^p g\, V \beta_p \beta_{n-p-1} \big]\ , \label{action2}
 \end{align}
while the gauge transformations read
 \begin{gather}
  \delta_\epsilon V = d \epsilon\ ,\quad \delta_\epsilon A_p = g\,
	\epsilon \beta_p\ ,\quad \delta_\epsilon A_{n-p-1} = g\,
	\epsilon \beta_{n-p-1} \notag \\
  \delta_\epsilon \beta_p = \delta_\epsilon \beta_{n-p-1} =
	\delta_\epsilon G_{\mu\nu} = 0\ . \label{trafo2}
 \end{gather}
The equations of motion for the auxiliary fields are coupled in exactly
the same way as the equations that determine the covariant derivatives
of $A_p$ and $A_{n-p-1}$,
 \begin{align}
  - (-)^{p(n-1)} \dual \beta_p = d A_{n-p-1} - g\, V \beta_{n-p-1}
	\notag \\
  - (-)^p \dual \beta_{n-p-1} = d A_p - g\, V \beta_p\ ,
 \end{align}
and the solutions are thus
 \begin{equation}
  \beta_p = \Delta' A_p\ ,\quad \beta_{n-p-1} = \Delta' A_{n-p-1}\ .
 \end{equation}

The action \eqref{action2} may also be used to derive a dual version of
our model. To that end one solves the equations of motion for $A_p$ and
$A_{n-p-1}$ through $\beta_p=dA_{p-1}$ and $\beta_{n-p-1}=dA_{n-p-2}$
and inserts the solution back into the action. The interaction vertex
$V\beta_p\beta_{n-p-1}$ then turns into a Chern-Simons term.
\medskip

\emph{Remark.}
If $n=1+4k$ and $p=2k$, one may identify $A_p$ with $A_{n-p-1}$. For
instance, in the case $n=5$, $p=2$ one then gets a 5-dimensional theory
involving only one 2-form gauge field besides $V$ and the gravitational
field.

\section*{Generalizations} 

\subsection*{\normalsize Coupling to Gravitational Chern-Simons Forms} %

We shall now discuss how the basic model introduced above can be
generalized by including couplings of $p$-form gauge potentials to
gravitational Chern-Simons forms.

We denote a gravitational Chern-Simons $p$-form by $q_p$, where
$p=3,7,\dots$, or $p=n-1$ in $n=2k$ dimensional spacetime (the latter
case corresponds to the Euler density in even dimensional spacetime).
So, for instance, in 5-dimensional spacetime there is only one
gravitational Chern-Simons form, $q_3$, which satisfies $d q_3
={R_\rho}^\sigma {R_\sigma}^\rho$, where ${R_\rho}^\sigma=
\frac{1}{2} dx^\mu dx^\nu {R_{\mu\nu\rho}}^\sigma$ (explicitly one has
$q_3={\Gamma_\mu}^\nu {R_\nu}^\mu - \frac{1}{3} {\Gamma_\mu}^\nu
{\Gamma_\nu}^\rho {\Gamma_\rho}^\mu$, where ${\Gamma_\rho}^\sigma=
dx^\mu {\Gamma_{\mu\rho}}^\sigma$). In 4-dimensional spacetime there are
two independent gravitational Chern-Simons 3-forms, $q_3$ and $q'_3$,
satisfying $d q_3={R_\rho}^\sigma {R_\sigma}^\rho$ and $d q'_3=
\varepsilon_{\mu\nu\rho\sigma}R^{\mu\nu}R^{\rho\sigma}$
respectively.

In a first order formulation, a gravitational Chern-Simons $(p+1)$-form
can be coupled to a $p$-form gauge potential through the following
extension of the action \eqref{action2},
 \begin{align}
  S = \half \int \big[ & d V \dual d V + (-)^{p(n-1)}\, \beta_p \dual
	\beta_p + (-)^{(p+1)(n-1)}\, \beta_{n-p-1} \dual \beta_{n-p-1}
	\label{action22} \\[-2pt]
  & + 2 \beta_p\, d A_{n-p-1} - 2 (-)^{n(p+1)} \beta_{n-p-1}\, (d
	A_p + q_{p+1}) - 2 (-)^p g\, V \beta_p \beta_{n-p-1} \big]\ .
	\notag
 \end{align}
This action is invariant under the gauge transformations \eqref{trafo1}
and \eqref{trafo2}, and the following modification of the spacetime
diffeomorphisms,
 \begin{align}
  \delta_\xi \Phi & = \mathcal{L}_\xi \Phi \quad \text{for}\ \Phi \in
	\{ G_{\mu\nu}, \beta_p, \beta_{n-p-1}, A_{n-p-1} \} \notag \\
  \delta_\xi A_p & = \mathcal{L}_\xi A_p - r_p\ , \label{trafo3}
 \end{align}
where $\mathcal{L}_\xi$ is the standard Lie derivative along the vector
field $\xi^\mu$ (with $\mathcal{L}_\xi \beta_p = (1/p!) \cdot
dx^{\mu_1}\!\dots dx^{\mu_p} \mathcal{L}_\xi \beta_{\mu_1\dots\mu_p}$
etc.), and $r_p$ is the $p$-form whose exterior derivative is the
non-covariant part of $\delta_\xi q_{p+1}$,
 \begin{equation}
  \delta_\xi q_{p+1} = \mathcal{L}_\xi q_{p+1} + d r_p\ .
 \end{equation}
So for $q_3$ as above, one gets $r_2=\partial_\mu \xi^\nu d
{\Gamma_\nu}^\mu$, for instance. As usual, the combination $d
A_p+q_{p+1}$ transforms covariantly under $\delta_\xi$,
$\delta_\xi(d A_p+q_{p+1})=\mathcal{L}_\xi (d A_p+q_{p+1})$, and
therefore the action \eqref{action22} is indeed invariant under
$\delta_\xi$.

{}From the above formulae one easily infers the second order version
of the action and the gauge transformations in presence of
Chern-Simons couplings. It is simply obtained from \eqref{deA} and
\eqref{S} by substituting $dA_p+q_{p+1}$ for $dA_p$ everywhere.
Note that this results in the presence of the gravitational
Chern-Simons form $q_{p+1}$ both in $\delta_\epsilon A_p$ and
$\delta_\epsilon A_{n-p-1}$.

\subsection*{\normalsize Dimensional Reduction} 

We shall now perform a dimensional reduction from $n$ to $n-1$
dimensions. This gives rise to couplings to additional gauge fields and
scalars. We shall discuss explicitly only the dimensional reduction of
the basic model. The extension to more involved models, such as those
with additional Chern-Simons couplings, is straightforward.

We denote the coordinates by
 \begin{equation}
  x^\mu = (x^a, x^n)\ ,\quad a = 1, \dots, n-1\ ,
 \end{equation}
and take all fields to be constant along the $x^n$-direction.
Then the metric decomposes in the usual manner,
 \begin{equation}
  G_{\mu\nu}\, dx^\mu \otimes dx^\nu = g_{ab}\, dx^a \otimes dx^b +
  \e^{2\varphi} (W + dx^n) \otimes (W + dx^n)\ ,
 \end{equation}
where $W = dx^a W_a$ is a 1-form in $n-1$ dimensions.

Upon dimensional reduction, $A_p$ gives rise to a $p$-form $\hat{A}_p$
and a $(p-1)$-form $\hat{A}_{p-1}$, while $A_{n-p-1}$ decomposes into
an $(n-p-1)$-form $\hat{A}_{n-p-1}$ and an $(n-p-2)$-form
$\hat{A}_{n-p-2}$. The connection $V$ introduces in addition to a
1-form $\hat{V}$ a scalar field $\phi$,
 \begin{equation}
  A_p = \hat{A}_p + \hat{A}_{p-1}\, dx^n\ ,\quad A_{n-p-1} =
  \hat{A}_{n-p-1} + \hat{A}_{n-p-2}\, dx^n\ ,\quad V = \hat{V}
  + \phi\, dx^n\ .
 \end{equation}
In the following, one should keep in mind that the descendants
$\hat{A}_p$, $\hat{A}_{n-p-1}$ and $\hat{V}$ all transform non-trivially
under the abelian gauge transformation associated with $W$, while
$\hat{A}_{p-1}$, $\hat{A}_{n-p-2}$ and $\phi$ are invariant,
 \begin{gather}
  \delta_\lambda W = d \lambda\ ,\quad \delta_\lambda \hat{A}_p
	= \hat{A}_{p-1}\, d \lambda\ ,\quad \delta_\lambda
	\hat{A}_{n-p-1} = \hat{A}_{n-p-2}\, d \lambda\ ,\quad
	\delta_\lambda \hat{V} = \phi\, d \lambda \notag \\
  \delta_\lambda \hat{A}_{p-1} = \delta_\lambda \hat{A}_{n-p-2} =
	\delta_\lambda \phi = 0\ .
 \end{gather}
$\delta_\lambda$ originates from a general coordinate transformation
in the $n$th direction.

When decomposing the dual of a $p$-form $\w_p$, we make use of the
relation
 \begin{equation*}
  \dual \w_p = \e^{-\varphi} \hstar \hw_{p-1} + (-)^p\, \e^\varphi
  \hstar\, (\hw_p - \hw_{p-1} W)\ (W + dx^n)\ ,
 \end{equation*}
where the Hodge star $\hstar$ in $n-1$ dimensions involves the
reduced Levi-Civita tensor $\hat{\ep}_{a_1 \dots a_{n-1}} =
\e^{-\varphi}\, \ep_{a_1 \dots a_{n-1}n}$ and indices are raised with
$g^{ab}$.

There are now two ways of determining the covariant
derivatives and gauge transformations of the $\hat{A}$-fields, which
turn out to give the same results. One may either start from the
first order formulation and solve the dimensionally reduced equations
of motion for the auxiliary fields, or one may directly decompose the
equations \eqref{DAp} and \eqref{DAn-p-1}. Let us follow the latter
approach: we define a covariant derivative $\hat{D}$ by the relations
 \begin{equation}
  D A_p = \hat{D} \hat{A}_p + \hat{D} \hat{A}_{p-1}\, dx^n\ ,\quad
  D A_{n-p-1} = \hat{D} \hat{A}_{n-p-1} + \hat{D} \hat{A}_{n-p-2}\,
  dx^n\ .
 \end{equation}
The decomposition of the left-hand sides is straightforward, and by
comparison of the terms with and without $dx^n$ one derives the
action of $\hat{D}$ on the four gauge fields. Similarly to the gauge
fields themselves, the descendants $\hat{D}\hat{A}_p$ and $\hat{D}
\hat{A}_{n-p-1}$ transform non-covariantly under $\delta_\lambda$. It
is convenient to use $\delta_\lambda$-invariant generalized field
strengths instead, analogous to those appearing in Kaluza-Klein
supergravity models (see e.g.\ \cite{DNP}),
 \begin{equation} \label{F}
  \cF_{p+1} = \hat{D} \hat{A}_p - \hat{D} \hat{A}_{p-1}\, W\ ,\quad
  \cF_{n-p} = \hat{D} \hat{A}_{n-p-1} - \hat{D} \hat{A}_{n-p-2}\, W\ .
 \end{equation}
Then also the connection $\hat{V}$ appears always in a
$\delta_\lambda$-invariant combination, which we denote by
 \begin{equation} \label{U}
  U = \hat{V} - \phi W\ .
 \end{equation}

Since the original $n$-dimensional fields do not appear anymore, we
shall omit all hats in the following. Explicitly, one finds
 \begin{align}
  \cF_{p+1} & = d A_p - d A_{p-1}\, W - g E^{-1} U \big[ \e^{-\varphi}
	\dual d A_{n-p-2} + g\, i_U (d A_p - d A_{p-1}\, W) \notag \\
  & \tab + (-)^p g\, \e^{-2\varphi} \phi\, d A_{p-1} \big] \notag
	\\[4pt]
  \cF_{n-p} & = d A_{n-p-1} - d A_{n-p-2}\, W - g E^{-1} U \big[ -
	(-)^{n(p+1)} \e^{-\varphi} \dual d A_{p-1} \notag \\
  & \tab + g\, i_U (d A_{n-p-1} - d A_{n-p-2}\, W) - (-)^{n-p} g\,
	\e^{-2\varphi} \phi\, d A_{n-p-2} \big]\ ,
 \end{align}
and
 \begin{align}
  D A_{p-1} & = d A_{p-1} - g E^{-1} U \big[ (-)^{n-p} \e^\varphi
	\dual (d A_{n-p-1} - d A_{n-p-2}\, W) + g\, i_U d A_{p-1} \big]
	\notag \\
  & \tab - g E^{-1} \phi \big[ (-)^p \e^{-\varphi} \dual d A_{n-p-2}
	+ (-)^p g\, i_U (d A_p - d A_{p-1}\, W) \notag \\
  & \tab + g\, \e^{-2\varphi} \phi\, d A_{p-1} \big] \notag \\[4pt]
  D A_{n-p-2} & = d A_{n-p-2} - g E^{-1} U \big[ - (-)^{(p+1)(n-1)}
	\e^\varphi \dual (d A_p - d A_{p-1}\, W) + g\, i_U d A_{n-p-2}
	\big] \notag \\
  & \tab - g E^{-1} \phi \big[ (-)^{p(n-1)} \e^{-\varphi} \dual d
	A_{p-1} - (-)^{n-p} g\, i_U (d A_{n-p-1} - d A_{n-p-2}\, W)
	\notag \\
  & \tab + g\, \e^{-2\varphi} \phi\, d A_{n-p-2} \big]\ ,
 \end{align}
where now
 \begin{equation}
  E = 1 + g^2 U^a U_a + g^2 \e^{-2\varphi} \phi^2\ .
 \end{equation}

The above expressions enter the gauge transformations $\delta_\epsilon$
of the forms $A_p$ etc., which are found to read
 \begin{align}
  \delta_\epsilon A_p & = g\, \epsilon \big[ \e^{-\varphi} \dual
	D A_{n-p-2} + (-)^{n-p} \e^\varphi \dual \cF_{n-p} W \big]
	\notag \\[2pt]
  \delta_\epsilon A_{p-1} & = (-)^{n-p} g\, \epsilon\, \e^\varphi \dual
	\cF_{n-p} \notag \\[2pt]
  \delta_\epsilon A_{n-p-1} & = - (-)^{n(p+1)} g\, \epsilon \big[
	\e^{-\varphi} \dual D A_{p-1} - (-)^p \e^\varphi \dual
	\cF_{p+1} W \big] \notag \\[2pt]
  \delta_\epsilon A_{n-p-2} & = - (-)^{(p+1)(n-1)} g\, \epsilon\,
	\e^\varphi \dual \cF_{p+1}\ ,
 \end{align}
while those of the remaining fields are simply
 \begin{equation}
  \delta_\epsilon U = d \epsilon\ ,\quad \delta_\epsilon \phi =
  \delta_\epsilon W = \delta_\epsilon \varphi = \delta_\epsilon\,
  g_{ab} = 0\ .
 \end{equation}

We observe that $DA_{p-1}$ and $DA_{n-p-2}$ contain terms which are
accompanied by the scalar $\phi$ rather than the connection $U$. One may
introduce ``minimal'' covariant derivatives $\cD$ of $A_{p-1}$ and 
$A_{n-p-2}$ by substracting appropriate covariant terms from
$DA_{p-1}$ and $DA_{n-p-2}$,
 \begin{align}
  \cD A_{p-1} & = D A_{p-1} + (-)^p g\, \e^{-\varphi} \phi \dual
	D A_{n-p-2} \notag \\
  \cD A_{n-p-2} & = D A_{n-p-2} + (-)^{p(n-1)} g\, \e^{-\varphi} \phi
	\dual D A_{p-1}\ . \label{Dmin}
 \end{align}
This results in the following simpler expressions:
 \begin{align}
  \cD A_{p-1} & = d A_{p-1} - g E^{-1} U \big[ (-)^{n-p} \cE\,
	\e^\varphi \dual (d A_{n-p-1} - d A_{n-p-2}\, W) + g\,
	i_U d A_{p-1} \notag \\
  & \tab - (-)^p g^2 \e^{-\varphi} \phi\, i_U \dual d A_{n-p-2} \big]
	\notag \\[4pt]
  \cD A_{n-p-2} & = d A_{n-p-2} - g E^{-1} U \big[ - (-)^{(p+1)(n-1)}
	\cE\, \e^\varphi \dual (d A_p - d A_{p-1}\, W) \notag \\
  & \tab + g\, i_U d A_{n-p-2} - (-)^{p(n-1)} g^2 \phi\, i_U \dual
	d A_{p-1} \big]\ ,
 \end{align}
where $\cE$ is a function of the scalar fields only,
 \begin{equation}
  \cE = 1 + g^2 \e^{-2\varphi} \phi^2\ .
 \end{equation}
Equations \eqref{Dmin} can be inverted to express $DA_{p-1}$ and
$DA_{n-p-2}$ in terms of the minimal covariant derivatives,
 \begin{align}
  D A_{p-1} & = \cE^{-1} \big[ \cD A_{p-1} - (-)^p g\, \e^{-\varphi}
	\phi \dual \cD A_{n-p-2} \big] \notag \\[4pt]
  D A_{n-p-2} & = \cE^{-1} \big[ \cD A_{n-p-2} - (-)^{p(n-1)} g\,
	\e^{-\varphi} \phi \dual \cD A_{p-1} \big]\ . \label{DDmin}
 \end{align}

Finally, we obtain the gauge invariant action in $(n-1)$-dimensional
spacetime by reduction of equation\ \eqref{S}. Using the equations
\eqref{F}, \eqref{U} and \eqref{DDmin}, it can be written entirely in
terms of $\delta_\lambda$-invariant as well as
$\delta_\epsilon$-covariant expressions,
 \begin{align}
  S = \half \int \Big[\, & \e^\varphi (dU + \phi\, dW) \dual (dU +
	\phi\, dW) + (-)^n \e^{-\varphi} d\phi \dual d\phi \notag
	\\[-2pt]
  & + (-)^{n(p+1)}\, \e^\varphi \big( \cF_{p+1} \dual \cF_{p+1} +
	\cF_{n-p} \dual \cF_{n-p} \big) \notag \\[2pt]
  & + (-)^{np}\, \cE^{-1} \e^{-\varphi} \big( \cD A_{p-1} \dual
	\cD A_{p-1} + \cD A_{n-p-2} \dual \cD A_{n-p-2} \big)
	\notag \\[2pt]
  & - 2g\, \cE^{-1} U \big( \dual \cD A_{p-1} \dual \cF_{n-p} -
	(-)^{n(p+1)} \dual \cD A_{n-p-2} \dual \cF_{p+1} \big)
	\notag \\[2pt]
  & - 2 (-)^{p(n-1)} g^2 \cE^{-1} \e^{-\varphi} \phi\, U \big(
	\cD A_{n-p-2} \dual \cF_{n-p} - (-)^n \cD A_{p-1} \dual
	\cF_{p+1} \big) \notag \\
  & + 2 (-)^p g\, \cE^{-1} \e^{-2\varphi} \phi\, \cD A_{p-1}
	\cD A_{n-p-2}\, \Big]\ . \label{DRaction}
 \end{align}

As compared to \eqref{S}, the new feature of the action \eqref{DRaction}
is the coupling of $\phi$, $\varphi$ and $W$. Indeed, if we set
these fields to zero, \eqref{DRaction} reduces to an action of the form
\eqref{S} in $n'=n-1$ dimensions, for two pairs of exterior form gauge
fields with form degrees $(p,n'-p-1)$ and $(p-1,n'-p)$: one gets
\eqref{S} with $X=X_1+X_2$ and $K=K_1+K_2$, where the subscripts 1 and 2
refer to the first and second pair respectively ($X_i$ and $K_i$ match
their counterparts in \eqref{X} and \eqref{K} up to sign factors which
can be removed by trivial field redefinitions $A\rightarrow\pm A$).
Clearly, one may iterate the dimensional reduction procedure to derive
more involved models of the above type. Again, after setting to zero
scalar and vector fields coming from $G_{\mu\nu}$ and $V$, the resulting
actions reduce to the form \eqref{S} for more pairs of exterior form
gauge fields.

\section*{Comments} 

To understand the structure of the models, it is useful to keep in mind
that they are consistent deformations of the standard Einstein-Maxwell
action for exterior form gauge fields, plus kinetic terms for the
additional fields in the models obtained by dimensional reduction. The
non-polynomial structure and the non-triviality of the interactions can
be traced to the properties of the global symmetry \eqref{DeA} that is
gauged.

Namely, to first order in the coupling constant $g$, the deformed
Einstein-Maxwell action \eqref{S} reads $S_1=(-)^{n-p}g\int VJ$, where
$J$ is the Noether current \eqref{current}. Hence, $S_1$ is nothing but
the standard Noether coupling of the gauge field $V$ to the conserved
current of the global symmetry \eqref{DeA}. The corresponding first
order deformation of the gauge symmetry involves the global symmetry
\eqref{DeA} itself: it is just $\delta_1 = g\epsilon \Delta$ (i.e.
$\delta_1 A_p = g\epsilon \dual dA_{n-p-1}$, $\delta_1 A_{n-p-1} =
- (-)^{n(p+1)} g\epsilon \dual dA_p)$. It can be readily checked
that $\delta_1 S_1$ does not vanish, due to the fact that $J$ is not
invariant under $\epsilon\Delta$. Rather, one has $\delta_1 S_1\approx
-\delta_0 S_2$, where $\approx$ denotes weak equality
(= ``on-shell equality''), i.e. equality up to terms that involve the
left hand sides of the equations of motion. As a consequence,
consistency of the interactions at second order makes it necessary to
introduce second order deformations both of the action and of the gauge
symmetry  (this can be rigorously proved by BRST cohomological
arguments along the lines of \cite{BH}). This extends to all higher
orders and leads to the non-polynomial structure of the interactions and
gauge transformations.

In fact the situation is somewhat similar to the coupling of the
gravitational field to matter in standard gravity. Indeed, viewed as a
deformation using $G_{\mu\nu}=\eta_{\mu\nu} + m_\mathrm{pl}^{-1}
h_{\mu\nu}$, the first order deformation couples $h_{\mu\nu}$ to the
energy momentum tensor $T^{\mu\nu}$ in flat spacetime. $T^{\mu\nu}$ is
the Noether current of translations and not invariant under the
generators of translations given by the spacetime derivatives. This
leads to interactions that are non-polynomial in $h_{\mu\nu}$ and in
the coupling constant $m_\mathrm{pl}^{-1}$ which are compactly
constructed by means of the familiar tensor calculus.

The non-standard coupling of the $p$-form gauge potentials to scalar
fields present in the dimensionally reduced models can be understood
analogously from the point of view of consistent deformations. If we
specialize these models by setting $W$ and $\varphi$ to zero, the first
order coupling of the scalar field $\phi$ is just $g\int\phi\, dA_{p-1}
dA_{n-p-2}$ (up to a factor), i.e. it couples $\phi$ to the topological
density $dA_{p-1} dA_{n-p-2}$. Again, the latter is not invariant under
$\delta_1$. This enforces further interaction terms of higher order,
with higher powers of $\phi$, and eventually non-polynomial interactions
of $\phi$. Analogous statements apply in presence of $W$ and $\varphi$,
where further interactions are present. One may disentangle the various
couplings in these interactions by introducing further coupling
constants $g_1$ and $g_2$ through the rescalings $W\rightarrow g_1W$,
$\varphi\rightarrow g_2\varphi$.

Hence, though one might have suspected that the non-polynomial
interactions of the scalar fields (and of $W$) in the above models
are just an artefact of the dimensional reduction, such interactions are
actually to be expected in gauge theories of the type studied here,
whether or not the models can be obtained by dimensional reduction. In
fact, the reader may check that non-polynomial interactions of scalar
fields similar to those found here are present in all supersymmetric
models constructed in \cite{CdWFKST,VT,BT}.

Finally we comment briefly on the non-triviality of the deformations
constructed here. Deformations of actions and gauge transformations are
called trivial if they can be removed by mere field redefinitions. The
non-triviality of the deformations constructed here traces to the
non-triviality of the global symmetry $\Delta$ and its current $J$
(again, this can be shown by BRST cohomological arguments).

In general, the non-triviality of conserved currents is concisely
formulated in the so-called characteristic cohomology (see e.g.\
\cite{char}) related to the equations of motion. This cohomology is
defined through the cocycle condition $d\w_p\approx 0$ and the
coboundary condition $\w_p\approx d\w_{p-1}$, where $\w_p$
and $\w_{p-1}$ are local forms constructed of the fields and their
derivatives (of first or higher order) and $\approx$ denotes weak
equality as described above. In our case $J$ is indeed a non-trivial
cocycle of the characteristic cohomology related to the Einstein-Maxwell
equations: it is conserved, $dJ\approx 0$, and non-trivial since
there is no local $(n-2)$-form $\w_{n-2}$ such that $J\approx d
\w_{n-2}$.

This implies already that $\Delta$ is non-trivial too, because a global
symmetry is trivial (= weakly equal to a gauge transformation) if and
only if the corresponding Noether current is trivial in the
characteristic cohomology \cite{BBH} (when global subtleties are absent
or irrelevant). In fact, in our case the non-triviality of $\Delta$
traces also directly to the characteristic cohomology: $\dual dA_{p}$
and $\dual dA_{n-p-1}$ are non-trivial in that cohomology (cf.\
\cite{HKS}) and thus $\Delta A_p$ and $\Delta A_{n-p-1}$ are not weakly
equal to a gauge transformation.

\emph{Remark.}
The non-triviality of $J$, $\dual dA_{p}$ and $\dual dA_{n-p-1}$ in the
characteristic cohomology does not contradict the fact that these
quantities are exact (at least locally) when one evaluates them for a
specific solution to the equations of motion. In fact, by the ordinary
Poincar\'e lemma, they are locally exact for \emph{any} specific
solution to the equations of motion since they are weakly closed. The
latter statement applies of course to every cocycle of the
characteristic cohomology (except for the constant 0-forms), and thus in
particular to every Noether current, whether or not it is trivial. The
point is that a form $\w_p$ which is trivial in the characteristic
cohomology is weakly equal to $d\w_{p-1}$ with a \emph{definite}
$\w_{p-1}$, that involves the fields and their derivatives and does
not depend upon which solution to the equations of motion one considers.
\bigskip

\textbf{Acknowledgements} \\[6pt]
FB and UT wish to thank Joaquim Gomis for his kind hospitality at the
University of Barcelona. This work was supported by the Acciones
Integradas programme of the Deutscher Akademischer Austauschdienst and
the Ministerio de Educaci\'on y Cultura. FB was supported by the
Deutsche Forschungsgemeinschaft. JS is supported by a fellowship from
Comissionat per a Universitats i Recerca de la Generalitat de Catalunya.

\end{document}